\documentstyle[aps,prl,epsfig,preprint]{revtex}

\def\onebetwo{\,}
\def\twobeone{\,}

\begin{document}

\title{One dimensional SU(3) bosons with $\delta$ function interaction}
\author{You-Quan Li, Shi-Jian Gu and Zu-Jian Ying}
\address{Zhejiang Institute of Modern Physics,
 Zhejiang University, Hangzhou 310027, P.R. China}
\date{\today}
\maketitle

\begin{abstract}
In this paper we solve one dimensional SU(3) bosons with repulsive
$\delta$-function interaction by means of Bethe ansatz method. The
features of ground state and low-lying excited states are studied
by both numerical and analytic methods. We show that the ground
state is a SU(3) color ferromagnetic state. The configurations of
quantum numbers for the ground state are given explicitly. For
finite $N$ system the spectra of low-lying excitations and the
dispersion relations of four possible elementary particles (holon,
antiholon, $\sigma$-coloron and $\omega$-coloron) are obtained by
solving Bethe-ansatz equation numerically. The thermodynamic
equilibrium of the system at finite temperature is studied by
using the strategy of thermodynamic Bethe ansatz, a revised
Gaudin-Takahashi equation which is useful for numerical method are
given . The thermodynamic quantities, such as specific heat, are
obtain for some special cases. We find that the magnetic property
of the model in high temperature regime is dominated by Curie's
law: $\chi\propto 1/T$ and the system has Fermi-liquid like
specific heat in the strong coupling limit at low temperature.
\end{abstract}


\section{Introduction}

One of the main goals of theoretical physics during the past 40
years is to understand quantum systems involving many particles.
Once the interaction between those particles were taken into
account, the problem becomes complicated. Meanwhile, as long as
their interaction is not sufficient weak, the perturbative methods
that were powerful in many quantum mechanics text book become
unreliable. In one dimension, various non-perturbative methods
were proposed, among which the impact of exactly solvable
theoretical models are undeniable. Particles with
$\delta$-function interaction is a simple but interesting model.
Lieb and Liniger \cite{EHLieb63} first solved a Bose system under
periodic boundary condition in the  case of spin-$0$ or in the
absence of internal degree of freedom. The method they used is
nowadays referred as coordinate Bethe ansatz. The extension of
periodic boundary condition to the boundary condition of potential
well of infinite depth were made by Gaudin\cite{MGaudin71} and
Woynarivich\cite{FWoynarovich85}. The first attempt to develop the
method applied in \cite{EHLieb63} to deal with spin-1/2 fermions
was made by McGuire\cite{JBMcGuire65} who can, however, deal with
the case of only one spin-down and the other spins keep up.
Further step of considering two spin-down and others spin-up was
made by Flicker and Lieb\cite{MFlicker67}. Gaudin\cite{MGaudin67}
and Yang \cite{CNYang67} successfully solved the problem for
arbitrary number of spins in the state of spin-down. In Yang's
paper\cite{CNYang67}, the first non-trivial case of Yang-Baxter
equation was introduced. Actually, the strategy for general
multi-component systems was proposed in \cite{CNYang67} though the
explicit solution was give only for spin-1/2 particles (it is
Fermi system then).

The literature in \cite{CNYang67} was later extended by
Sutherland\cite{BSutherland68} to the called any irreducible
representation of permutation group. Actually, both Yang and
Sutherland adopted antisymmetric wave function under permutating
indistinguishable particles. Thus Yang solved the problem of
2-component fermions and Sutherland solved N-component fermions by
means of coordinate Bethe ansatz. As 2-component system is mostly
associated with ``spin-1/2" system which is conventionally
referred to Fermi system, the coordinate Bethe ansatz has not been
employed to 2-component Bose systems till recently \cite{YQLi01}.

Along with the developments of quantum inverse scattering method,
Kulish\cite{PPKulish81} discussed multi-component nonlinear
Schr\"odinger equation in terms of quantum inverse scattering
method (QISM) in order to re-derive the Bethe-ansatz equations of
Yang\cite{CNYang67} and Sutherland\cite{BSutherland68}. The author
explicitly formulated 2-component case and conjectured that
Sutherland's results would be obtained by repeating his procedure
$n-2$ times. Actually, it is not possible because the
author\cite{PPKulish81} adopted commutation (instead of
anti-commutation) relations but the system both Yang and
Sutherland considered is Fermi system. It is now clear that the
first quantization form of the system which Kulish considered
ought to be a system of SU(n) bosons with $\delta$-function
interaction. The QISM  was also employed to nonlinear
Schr\"odinger equation for graded matrix but it breaks the
Yang-Baxter relation at first \cite{PPKyulish80} that was noticed
and overcome later on \cite{FanPuZhao}.

Although the Bethe-ansatz equations for 2-component bosons was
early formulated by Kulish, the nature of ground state and the
property of low-lying excitations has never been exposed till the
paper of Li {\it et. al.}\cite{YQLi01}. We know that not only the
2-component Bose gas can be formed in magnetically trapped
$^{87}Rb$\cite{WilliamsEXP}, but also a 3-component Bose gas can
be produced in an optically trapped $^{23}Na$\cite{Ketterle}, it
will be valuable to study the model of 3-component Bose system. In
present paper, we study a system of three-component bosons with
SU(3) symmetry in one dimension. On the basis of Bethe-ansatz
equations, we discuss the ground state, low-lying excited states
and the thermodynamics of the system at finite temperature, and
try to obtain thermal coefficients for some special cases. Our
paper is organized as follows: In the following section we
introduce the model and the corresponding Bethe-ansatz equations
for charge rapidity and color rapidities. In Sec.
\ref{sec:ground}, we explicitly show that the ground state is a
color ferromagnetic state and how the quantum numbers in
Bethe-ansatz equations should be taken for the ground state. In
Sec. \ref{sec:low}, We study the low-lying excited states
extensively by analyzing the possible variations in the sequence
of quantum numbers. Numerical results of energy momentum spectra
for each excitation are given. Furthermore, the dispersion
relation of four possible elementary particles are obtained. In
Sec. \ref{sec:thermo} we discuss the general thermodynamics of the
system with the strategy of thermodynamic Bethe ansatz (TBA) which
was proposed by C.N. Yang and C.P. Yang when they study Bose gas
with $\delta$-function interaction in
one-dimension\cite{CNYang69}. In Sec. \ref{sec:spec} we discuss
the system for some special cases and obtain some analytical
results. In Sec. \ref{conclusions} a brief summary is given.

\section{The model and its Bethe-ansatz solution}\label{sec:model}

We consider interacting SU(3) Bose field in a one dimensional ring
of length $L$. The model Hamiltonian of the system reads,
\begin{equation}
{\cal H}_0=\int dx \left[ \sum_a\partial_x\psi_a^*
\partial_x\psi_a+\frac{c}{2}\sum_{a, b}\psi_a^*\psi_a
\psi_b^* \psi_b \right ], \label{eq:Hamiltonian}
\end{equation}
where natural unit is adopted for simplicity. Here $c$ is the
coupling constant and $a, b=1, 2, 3$ (we call colors hereafter)
denotes the three states that carry out the fundamental
representation of SU(3) group. The fields obey the following
commutation relations,
\begin{equation}
[\psi_a^*(x),\psi_b(y)]=\sum_n \delta_{ab}\delta(x-y-nL).
\end{equation}
In the terminology of group theory, the three states $|1\rangle$,
$|2\rangle$ and $|3\rangle$ are labeled by weight vectors, $(1/2,
0)$, $(-1/2, 1/2)$ and $(0, -1/2)$ respectively. The two $su(2)$
subalgebra in the $su(3)$ Lie algebra are $[T^+, T^-]=2T^z$ and
$[U^+, U^-]=2U^z$. With the help of those ``flipping'' operators,
$U^\pm$ and $T^\pm$, we can generate the three states from the
highest weight state $|1\rangle$, i.e.,
\[
 T^-|1\rangle=|2\rangle ,\,
 U^-|2\rangle=|3\rangle.
\]
With additional commutation relations defined by
\begin{eqnarray}
V^+=[T^+, U^+],\; V^-=[T^-, U^-], \nonumber
\end{eqnarray}
the Chevalley bases of the $su(3)$ Lie algebra consists of eight
generators $\{T^{\pm}, U^{\pm}, V^{\pm}, T^z, U^z \}$.

In the domain with $x_i\neq x_j$, the Hamiltonian
(\ref{eq:Hamiltonian}) reduces to the one for free bosons and its
eigenfunction is therefore just the superposition of plane waves.
When two particles collide with each other, a scattering process
occurs. The coordinate Bethe ansatz embodies that this process is
purely elastic, i.e., exchange of their momenta. So for a given
momentum $k=(k_1, k_2,\dots, k_N)$, the scattering momenta include
all permutations of the components of $k$. Thus for the case of
$N$ bosons, due to the Hamiltonian is invariant under the action
of the permutation group $S_N$, one can adopt the following
Bethe-ansatz wave function,
\begin{equation}
\Psi_a (x)=\sum_{P\in {S_N}}A_a(P,Q)e^{i(P k|Qx)}, \; x \in {\cal
C}(Q), \label{eq:BAWF}
\end{equation}
where $a=(a_1, a_2,\ldots,a_N)$, $a_j$ denotes the color label of
the $j$th particle; $Pk $ stands for the image of a given $ k:= (
k_1 , k_2, \cdots , k_N ) $ by a mapping $P\in S_N $ and the
coefficients $ A(P,Q)$ are functionals of $P$ and $Q$ where the
$Q$ denotes a permutation of the coordinates which defines a
region with $0<x_{Q_1}<x_{Q_2}<\dots <x_{Q_N}<L$. For a Bose
system, the wave function is supposed to be symmetric under any
permutation of both coordinates and color indices, i.e.
\begin{equation}
(\Pi^j\Psi)_a (x)=\Psi_a (x),
\label{eq:permutation}
\end{equation}
where $\Pi^j:\, \{a_1,...,a_j, a_{j+1},...\} \mapsto
\{a_1,...,a_{j+1}, a_j,...\}$ and $(\Pi^j \Psi)_a$ is well defined
by $\Psi_{\Pi^j a}(\Pi^j x)$. Furthermore, using the identity
$(Pk|\Pi^i Qx) = (\Pi^i Pk|Qx)$ and rearrangement theorem of group
theory, we have the following consequence from
(\ref{eq:permutation}):
\begin{equation}
A_a(P,\Pi^i Q)=A_{\Pi^j a}(\Pi^i P,Q).
\label{eq:symmetry}
\end{equation}

The $\delta$-function term in the Hamiltonian
(\ref{eq:Hamiltonian}) contributes a boundary condition across the
hyper-plane  $x_{Q_j}=x_{Q_{j+1}}$,
\twobeone
\begin{equation}
i\bigl((Pk)_j-(Pk)_{j+1}\bigr)\bigl[ A_a(P,\Pi^j Q)-A_{a}(\Pi^j P,
\Pi^j Q) -A_{a}(P, Q)+A_{a}(\Pi^j P, Q)]
  =2c[A_a(P, Q)+A_a(\Pi^j P, Q)],
\label{eq:BCCC}
\end{equation}
\onebetwo
By making use of  the relations (\ref{eq:symmetry}) and
(\ref{eq:BCCC}) together with the continuity condition, we can
obtain the following relation
\begin{equation}
A_{a}(\Pi^j P, Q)=\frac{i[(Pk)_j-(Pk)_{j+1}]{\cal P}^j+c}
 {i[(kP)_j-(Pk)_{j+1}]-c}A_a(P, Q),
\label{SM1}
\end{equation}
where ${\cal P}^j$ permutates the color labels of bosons located
at $x_{Q_j}$ and $x_{Q_{j+1}}$.

Applying the periodic boundary condition
$\Psi(\cdots,x_{Q_j},\cdots)=\Psi(\cdots,x_{Q_j}+L,\cdots)$ and
making use of the standard procedure of quantum inverse scattering
method\cite{BSutherland68,PPKulish81,LDFaddeev84}, one can obtain
the Bethe-ansatz equations
\begin{eqnarray}
e^{ik_jL}=&&-\prod^{N}_{l=1}\frac{k_j-k_l+ic}{k_j-k_l-ic}
   \prod_{\nu=1}^{M}\frac{k_j-\lambda_\nu-ic/2}{k_j-\lambda_\nu+ic/2},
    \nonumber\\
1=&&-\prod^{N}_{l=1}
    \frac{\lambda_\gamma-k_l-ic/2}{\lambda_\gamma-k_l+ic/2}
     \prod^{M}_{\nu=1}
      \frac{\lambda_\gamma-\lambda_\nu+ic}{\lambda_\gamma-\lambda_\nu-ic}
       \nonumber \\
       &&\times\prod^{M'}_{\alpha=1}
\frac{\lambda_\gamma-\mu_\alpha-ic/2}{\lambda_\gamma-\mu_\alpha+ic/2},
         \nonumber\\
1=&&-\prod^{M}_{\nu=1}
    \frac{\mu_\beta-\lambda_\nu-ic/2}{\mu_\beta-\lambda_\nu+ic/2}
     \prod^{M'}_{\alpha=1}
      \frac{\mu_\beta-\mu_\alpha+ic}{\mu_\beta-\mu_\alpha-ic}.
\label{eq:BAE3}
\end{eqnarray}
The $\lambda$ and $\mu$ are SU(3) color rapidities. There are
$M-M'$ particles in the state $|2\rangle$, $M'$ in $|3\rangle$ and
$N-M$ in $|1\rangle$.  We would like to mention here that the
state obtained above is the highest weight state among the
multiplet of SU(3) representation labeled by ($N/2+M'/2-M,
M/2-M'$). The other states in the multiplets can be generated by
iterate application of the flipping operators $T^-$ and $U^-$.

Taking logarithm of Eqs. (\ref{eq:BAE3}) we have secular equations,
\begin{eqnarray}
k_jL &=& 2\pi I_j +\sum_{l=1}^N \Theta_1(k_j-k_l)
 +\sum_{\nu=1}^M\Theta_{-1/2}(k_j-\lambda_\nu),
   \nonumber \\
2\pi J_\gamma&=&\sum_{l=1}^N\Theta_{-1/2}(\lambda_\gamma-k_l)
  +\sum_{\nu=1}^M\Theta_1(\lambda_\gamma-\lambda_\nu) \nonumber\\
   &&+\sum_{\alpha=1}^{M'}
   \Theta_{-1/2}(\lambda_\gamma-\mu_\alpha),
\nonumber \\
2\pi
{J'}_\beta&=&\sum_{\nu=1}^M\Theta_{-1/2}(\mu_\beta-\lambda_\nu)
   +\sum_{\alpha=1}^{M'}
   \Theta_1(\mu_\beta-\mu_\alpha),
\label{eq:logBAE3}
\end{eqnarray}
where $\Theta_n(x)=-2\tan^{-1}(x/nc)$. The quantum number $I_j$
for charge rapidity $k_j$ takes integer or half-integer depending
on whether $N-M$ is odd or even. The quantum number $J_\gamma$ and
${J'}_\beta$ for SU(3) color rapidities $\lambda_\gamma$ and
$\mu_\beta$ take integer or half-integer depending on whether
$N-M-M^\prime$ and $M-M^\prime$ is odd or even respectively. Once
all roots $\{k_j, \lambda_\gamma, \mu_\beta\}$ are solved from the
above equations (\ref{eq:logBAE3}) for a given set of quantum
numbers $\{I_j, J_\gamma, {J'}_\beta\}$, the energy and momentum
will be calculated by
\begin{eqnarray}
E=&&\sum_{j=1}^N k_j^2,\nonumber \\
p=&&\frac{2\pi}{L}\left [ \sum_{j=1}^N I_j-\sum_{\gamma=1}^M
J_\gamma-\sum_{\beta=1}^{M^\prime}{J'}_\beta \right ],
\label{eq:EP}
\end{eqnarray}
where the second equation of Eqs. (\ref{eq:EP}) is obtained
from Eqs. (\ref{eq:logBAE3}) directly.

For a state with real roots ($k, \lambda, \mu$), we may define the
distribution densities $\rho(k)$, $\sigma(\lambda)$ and
$\omega(\mu)$
\begin{eqnarray}
\rho(k_j)=&&1/L(k_{j+1}-k_j),\nonumber \\
\sigma(\lambda_\gamma)=&&1/L(\lambda_{\gamma+1}-\lambda_\gamma),\nonumber \\
\omega(\mu_\beta)=&&1/L(\mu_{\beta+1}-\mu_\beta). \label{eq:Den}
\end{eqnarray}
In terms of those densities the energy and momentum become
\begin{eqnarray}
E/L=&&\int k^2\rho(k)dk,\nonumber \\
p/L=&&\int k\rho(k)dk.
\end{eqnarray}
While $N$, $M$ and $M'$ are determined by
\begin{eqnarray}
N/L=&&\int\rho(k)dk,\nonumber \\
M/L=&&\int\sigma(\lambda)d\lambda,\nonumber\\
M'/L=&&\int\omega(\mu)d\mu.
\label{eq:density222}
\end{eqnarray}
As the SU(3) ``magnetic'' field is characterized by two parameters
$H_1$ and $H_2$, the Zeemann term is given by
\begin{eqnarray}
{\cal H}_{zee}&&=H_1(N-2M+M')/2 + H_2 (M-2M')/2\nonumber \\
&&=\frac{H_1 L}{2}\int\rho(k)dk
   +\frac{(H_2-2H_1)L}{2}\int\sigma(\lambda)d\lambda \nonumber \\
   &&+\frac{(H_1-2H_2)L}{2}\int\omega(\mu)d\mu.
\label{eq:Zee}
\end{eqnarray}

\section{The ground state}\label{sec:ground}
\begin{figure}
\setlength\epsfxsize{75mm}
\epsfbox{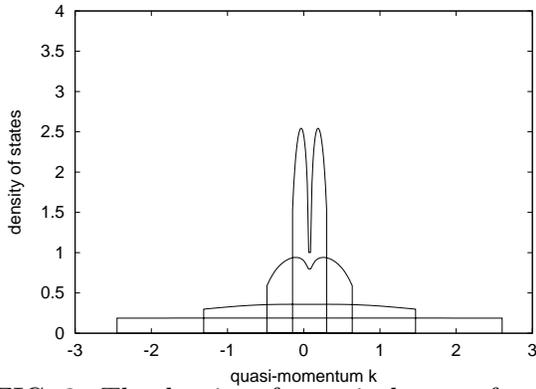}
\caption{The density of state in $k$-space
for the ground state in the
presence of one color rapidity by
choosing $J_1=0$. The distribution
changes from a histogram to a narrow
peak gradually for the coupling from
strong to weak. The figure is plotted
for $N=L=100$ and $c=10,1, 0.1, 0.01$.}
\label{FIGURE_RHOKM}
\end{figure}

It is easy to show that the first equation of Eqs.
(\ref{eq:logBAE3}) is a monotonously increasing function of $k_j$,
that is if $I_i<I_j$ we have $k_i<k_j$. So the configuration of
$\{I_j\}$ for the ground state is given by succussive integers or
half integers that is symmetrically arranged around zero, i.e.,
$I_{j+1}-I_j=1$. In order to observe the properties of
$\{J_\gamma, {J'}_\beta\}$, it is useful to investigate the Eqs.
(\ref{eq:logBAE3}) in the weak coupling limit $c\rightarrow 0$.
Due to $\Theta_{\pm n}(x)\rightarrow \mp\pi{\rm sgn}(x)$, Eqs.
(\ref{eq:logBAE3}) become

\begin{eqnarray}
2I_j=&&k_jL/\pi+\sum_{l=1}^N{\rm sgn}(k_j-k_l)
   -\sum_{\nu=1}^M{\rm sgn}(k_j-\lambda_\nu),\nonumber\\
2 J_\gamma=&&\sum_{l=1}^N{\rm sgn}(\lambda_\gamma-k_l)
   -\sum_{\nu=1}^M{\rm sgn}(\lambda_\gamma-\lambda_\nu)
   \nonumber \\
   &&
   +\sum_{\alpha=1}^{M'}{\rm sgn}(\lambda_\gamma-\mu_\alpha),\nonumber\\
2{J'}_\beta=&&\sum_{\nu=1}^M{\rm sgn}(\mu_\beta-\lambda_\nu)
   -\sum_{\alpha=1}^{M'}{\rm sgn}(\mu_\beta-\mu_\alpha).
\end{eqnarray}
We can choose the subscripts of the rapidities $k_j,
\lambda_\gamma, \mu_\beta$ in such a way that $I_j, J_\gamma,
{J'}_\beta$ are all ranged in an increasing order. Then we have
\twobeone
\begin{eqnarray}
2(I_{j+1}&&-I_j-1)=\frac{L}{\pi}(k_{j+1}-k_j)
 -\sum_{\nu=1}^M\bigl[{\rm sgn}(k_{j+1}-\lambda_\nu)
   -{\rm sgn}(k_j-\lambda_\nu)\bigr],
     \nonumber \\
2(J_{\gamma+1}&&-J_\gamma+1)=\sum_{l=1}^N
   \bigl[{\rm sgn}(\lambda_{\gamma+1}-k_l)
   -{\rm sgn}(\lambda_\gamma-k_l)\bigr]
    +\sum_{\alpha=1}^{M'}
     \bigl[{\rm sgn}(\lambda_{\gamma+1}-\mu_\alpha)
      -{\rm sgn}(\lambda_\gamma-\mu_\alpha)\bigr],
       \nonumber \\
2({J'}_{\beta+1}&&-{J'}_\beta+1)=\sum_{\nu=1}^M
   \bigl[{\rm sgn}(\mu_{\beta+1}-\lambda_\nu)
   -{\rm sgn}(\mu_\beta-\lambda_\nu)\bigr].
\label{eq:C0limit}
\end{eqnarray}
\onebetwo
Therefore, if ${J'}_{\beta+1}-{J'}_\beta=m$, there must exist
$m+1$ solutions of $\lambda_\nu$ satisfying
$\mu_\beta<\lambda_\nu<\mu_{\beta+1}$; and if
$J_{\gamma+1}-J_\gamma=n$, there must be $n+1$ solutions of $k_l$
and $\mu_\alpha$ satisfying $\lambda_\gamma<k_l,
\mu_\alpha<\lambda_{\gamma+1}$. So the existence of a
$\lambda_\nu$ between two $\mu$'s has a positive contribution to
the density of $\mu$ (\ref{eq:Den}), and vice versa for $\mu$ to
$\lambda$. However, from the first equation of Eqs.
(\ref{eq:C0limit}), for $I_{j+1}-I_j=n$, there will be
$k_{j+1}-k_j=2n\pi/L$ if there is a $\lambda_\gamma$ such that
$k_j<\lambda_\gamma<k_{j+1}$, otherwise $k_{j+1}-k_j=2(n-1)\pi/L$.
So a rapidity of $\lambda_\gamma$ always repels the $k$ rapidity
away from that value. As a result, an existing $\lambda_\gamma$
will suppress the density of state in $k$-space at the point
$k=\lambda_\gamma$. The weaker the coupling, the more magnificent
the effect will be (see Fig.\ref{FIGURE_RHOKM}). And for a given
set $\{I \}$, the more $\lambda$ rapidities there are, the higher
the energy is.

It is also useful to observe Eq. (\ref{eq:C0limit}) in the strong
coupling limit. We consider two cases: $M=0$ and $M=1$. For $M=0$,
the secular equation becomes
\begin{equation}
k_j L=2\pi I_j +\sum_{l=1}^N\Theta_1(k_j-k_l).
\end{equation}
and for $M=1$ we have
\begin{eqnarray}
k'_j L=2\pi I'_j+\sum_{l=1}^N\Theta_1(k'_j-k'_l)
 +\Theta_{-1/2}(k'_j-\lambda_1).
\end{eqnarray}
Here $I_j-I'_j=1/2$ due to $M$ changing from zero to one. As
$c\rightarrow \infty$, we have $\tan^{-1}(x/c)\sim x/c$. So the
above two equations become
\begin{eqnarray}
(k_{j+1}-k_j)L\left[1+\frac{2N}{Lc}\right]=&& 2\pi,\nonumber \\
(k'_{j+1}-k'_j)L\left[1+\frac{2(N-1)}{Lc}\right]=&&2\pi.
\end{eqnarray}
Whence the distribution is almost a histogram. Referring to Eq.
(\ref{eq:Den}) the value of the density distribution for $M=0$ is
larger than that for $M=1$, which makes the Fermi momentum for
later case to be larger than the former case so that to keep the
total number of particles being the same. Therefore the state of
$M=0$ has lower energy.

\begin{figure}
\setlength\epsfxsize{75mm}
\epsfbox{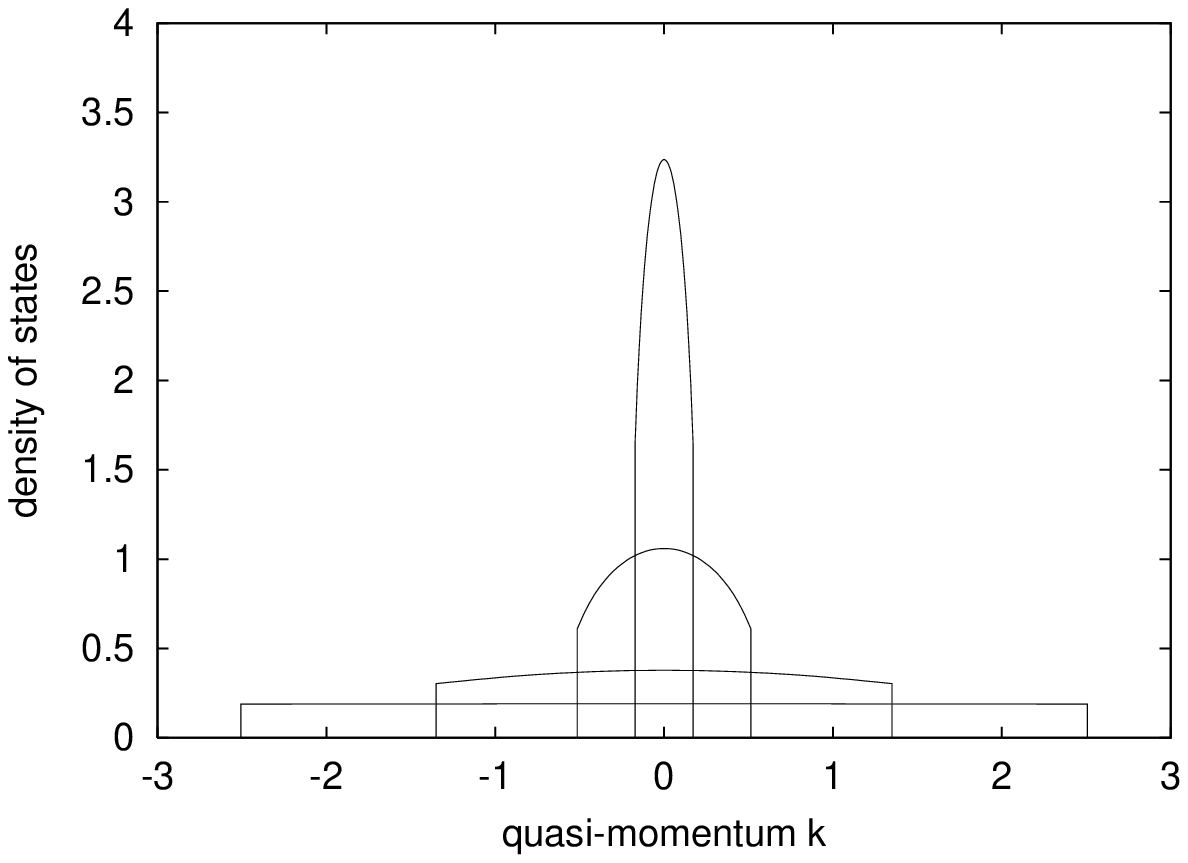}
\caption{The density of state in $k$-space
for the ground state. The distribution
changes from a histogram to a narrow peak
gradually for the coupling from
strong to weak. The figure is plotted
for $N=L=41$ and $c=10,1, 0.1, 0.01$.}
\label{FIGURE_RHOK}
\end{figure}

Differing from the SU(3) fermionic model\cite{BSutherland68} and a
toy model of quark cluster\cite{Koltun}, the ground state of SU(3)
bosonic model is no more a color singlet but a color ferromagnetic
state. The difference is due to the distinct permutation
symmetries.  For $N$ particles, the ground state is characterized
by a one-row $N$-column Young tableau $[ N ]$ whose quantum-number
configurations are
\begin{eqnarray}
\{I_j^0\}&&=\{-(N-1)/2,\cdots,(N-1)/2\} \nonumber \\
M&&=M'=0.
\label{eq:GN}
\end{eqnarray}
The density of states for the ground state is plotted in Fig.
\ref{FIGURE_RHOK} for various coupling with $L=N=41$.

\begin{figure}
\setlength\epsfxsize{75mm}
\epsfbox{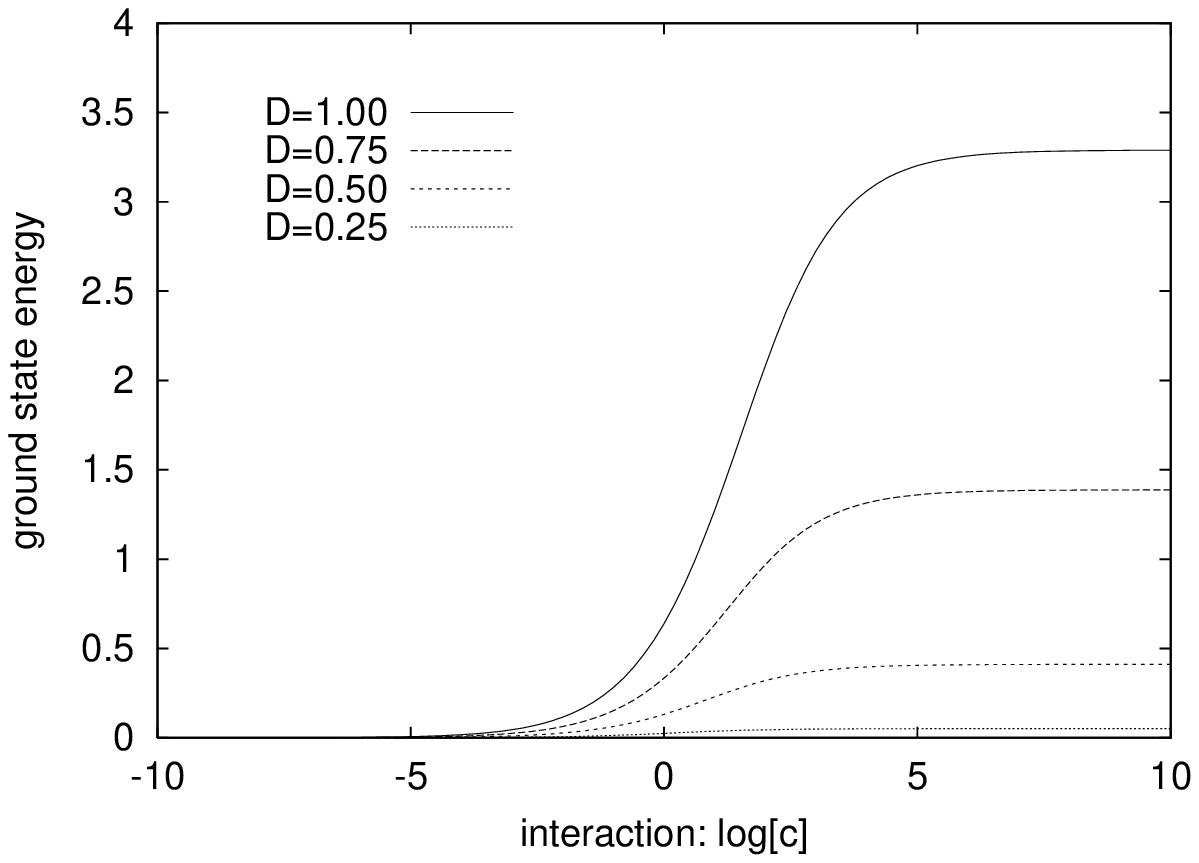}
\caption{The ground state energy $E/L$ versus the coupling constant $\ln c$
for different densities $D=1.0, 0.75, 0.5, 0.25$.}
\label{FIGURE_GE}
\end{figure}

In the thermodynamic limit, the density corresponding to the
configuration of quantum numbers of the ground state satisfies the
integral equation
\begin{equation}
\rho_0(k)=\frac{1}{2\pi}
 +\int_{-k_F}^{k_F}K_2(k-k^\prime)\rho_0(k^\prime)dk^\prime.
\label{eq:GRHO}
\end{equation}
Here $\rho_0(k), k_F$ are respectively the density and integration
limit for the ground state, and
\[
K_n(x)=\frac{1}{\pi}\frac{nc/2}{n^2c^2/4+x^2},
\]
The concentration is given by
\begin{equation}
D=N/L=\int_{-k_F}^{k_F}\rho_0(k)dk.
\label{eq:GN1}
\end{equation}
From Eq. (\ref{eq:GRHO}) and Eq. (\ref{eq:GN1}), we can determine
$\rho_0(k)$ and $k_F$. Here $k_F$ is a quasi-Fermi momentum
because the wave function vanishes for any $k_j=k_l$ $(j\neq l)$
as long as $c\neq 0$ even in Bose system which can be seen from
Eq.(\ref{SM1}). The energy can be calculated by
\begin{equation}
E_0/L=\int_{-k_F}^{k_F}k^2\rho_0(k)dk.
\label{eq:GroundSE}
\end{equation}
which is explicitly
$\displaystyle\frac{1}{3}\pi^2D^3(1-\frac{4}{c}D)$ in the strong
coupling limit. In the general case one needs to solve the
equations numerically. We show the ground state energy for
particle densities $D=1.0, 0.75, 0.5$ in Fig. \ref{FIGURE_GE}.

\section{Low-lying excited states}
\label{sec:low} The low-lying excited states are obtained by
varying the configuration $\{I_j, J_\gamma, {J'}_\beta\}$ from
that of the ground state.

\begin{figure}
\epsfclipoff
\fboxsep=0pt
\setlength{\unitlength}{0.8mm}
\begin{picture}(80,50)(0,0)
\linethickness{1pt}
\epsfysize=45mm
\put(-1,1){{\epsffile{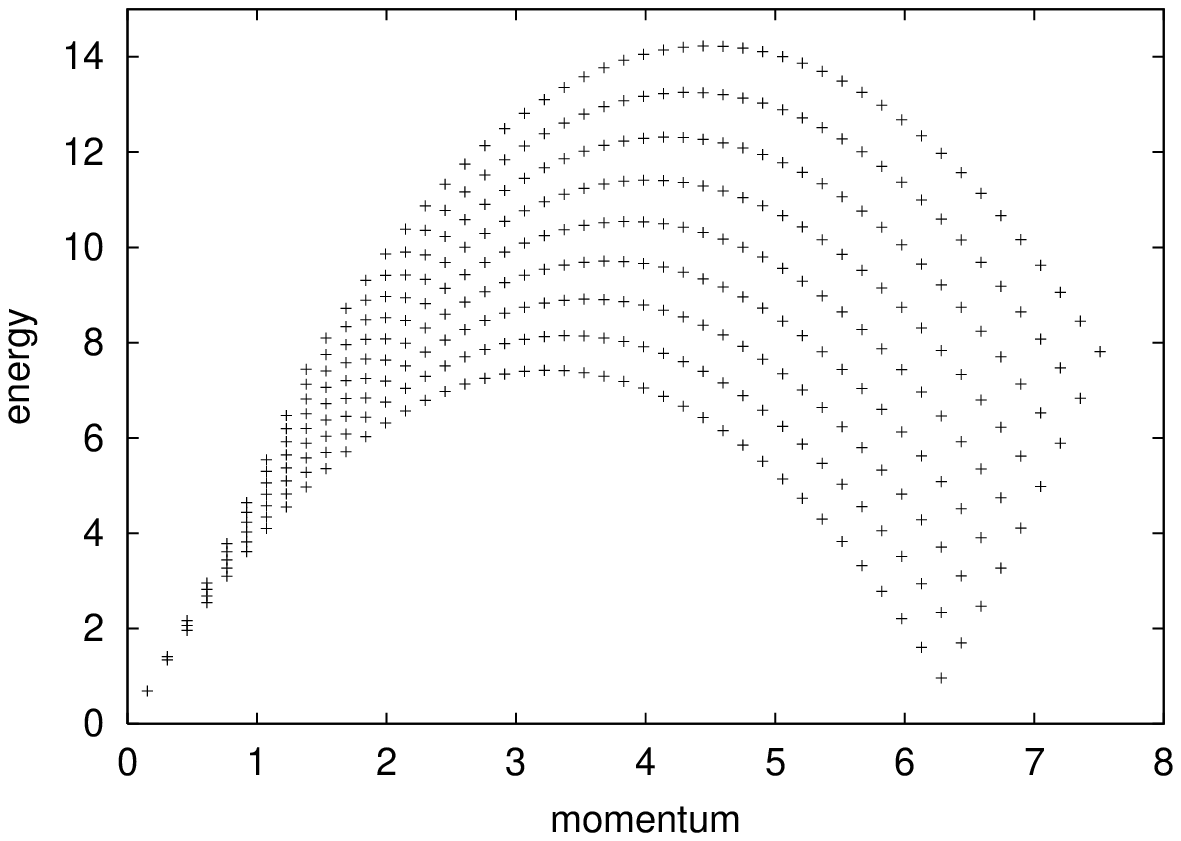}}}
\epsfysize=45mm
\put(90,1){{\epsffile{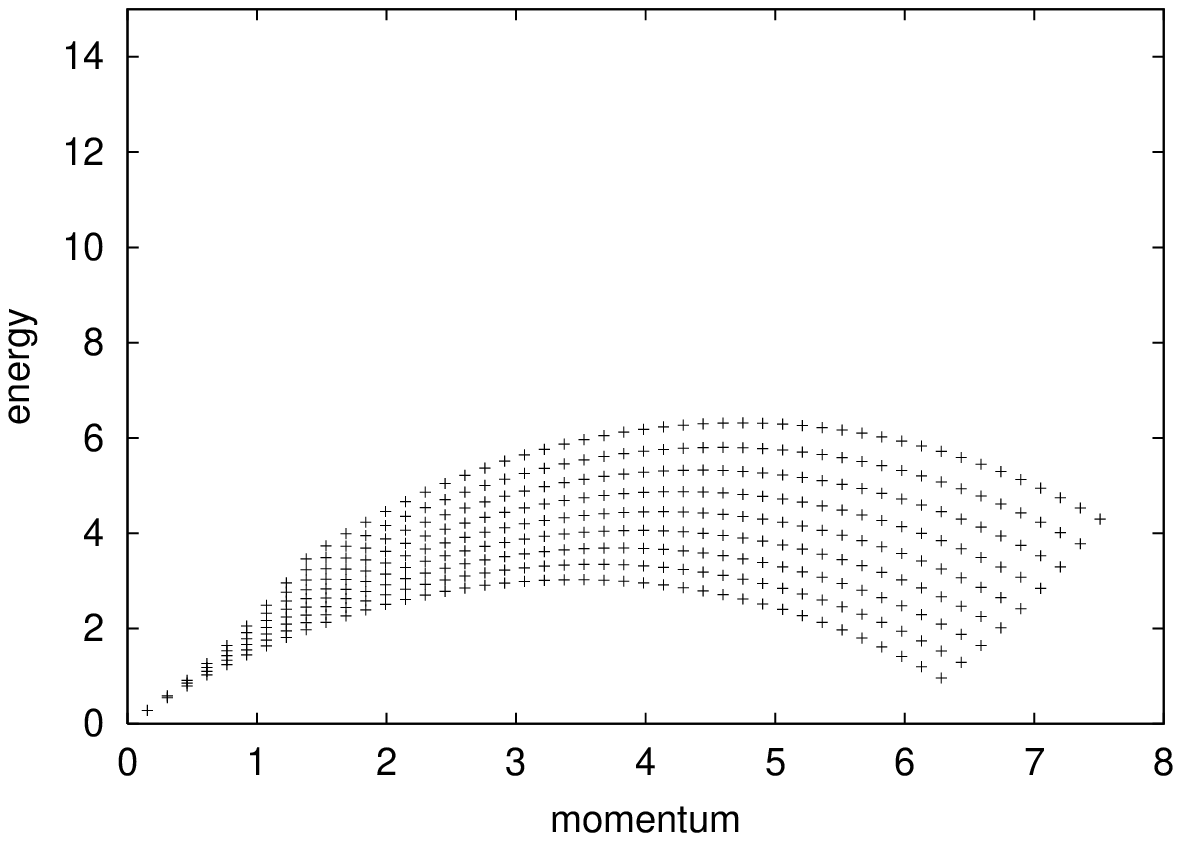}}}
\end{picture}
\vspace{0mm} \caption{The holon-antiholon excitation spectrum
calculated for $N=L=41$ and $c=10$ (left), $c=1$ (right).}
\label{FIGURE_PH}
\end{figure}

{\bf Holon-antiholon excitation}. The simplest case is to remove
one of $I$ from the configuration of the ground state and add a
new one outside the original sequence, i.e.,
$$
\{I_j\}=\{-(N-1)/2,...,n_1-1, n_1+1,...,(N-1)/2, I_n\},
$$
where $|I_n|>(N-1)/2$ and $M=M'=0$. We call this holon-antiholon
excitation which consists of a ``holon" created under Fermi
surface and an ``antiholon" created outside it. In Fig.
\ref{FIGURE_PH}, we plot the numerical results of energy-momentum
spectrum for a system with $L=N=41$ (the other part is just the
mirror image of the plotted part corresponding to the state with
$p\rightarrow -p$ coming from the negative $I_n$). From the
figure, we notice that there is a minimum in the excitation energy
at $p=2\pi$ due to the fact that both $I^0_1$ replaced by
$I_n=(N+1)/2$ and $I^0_N$ share the same energy, their momenta
difference, however, is $2\pi$. The overall structure of the
spectrum is not changed obviously between $c=1$ and $c=10$. For a
system of finite size, the gap of holon-antiholon excitations
opens. In the thermodynamic limit, however, it vanishes.

\begin{figure}
\setlength\epsfxsize{75mm}
\epsfbox{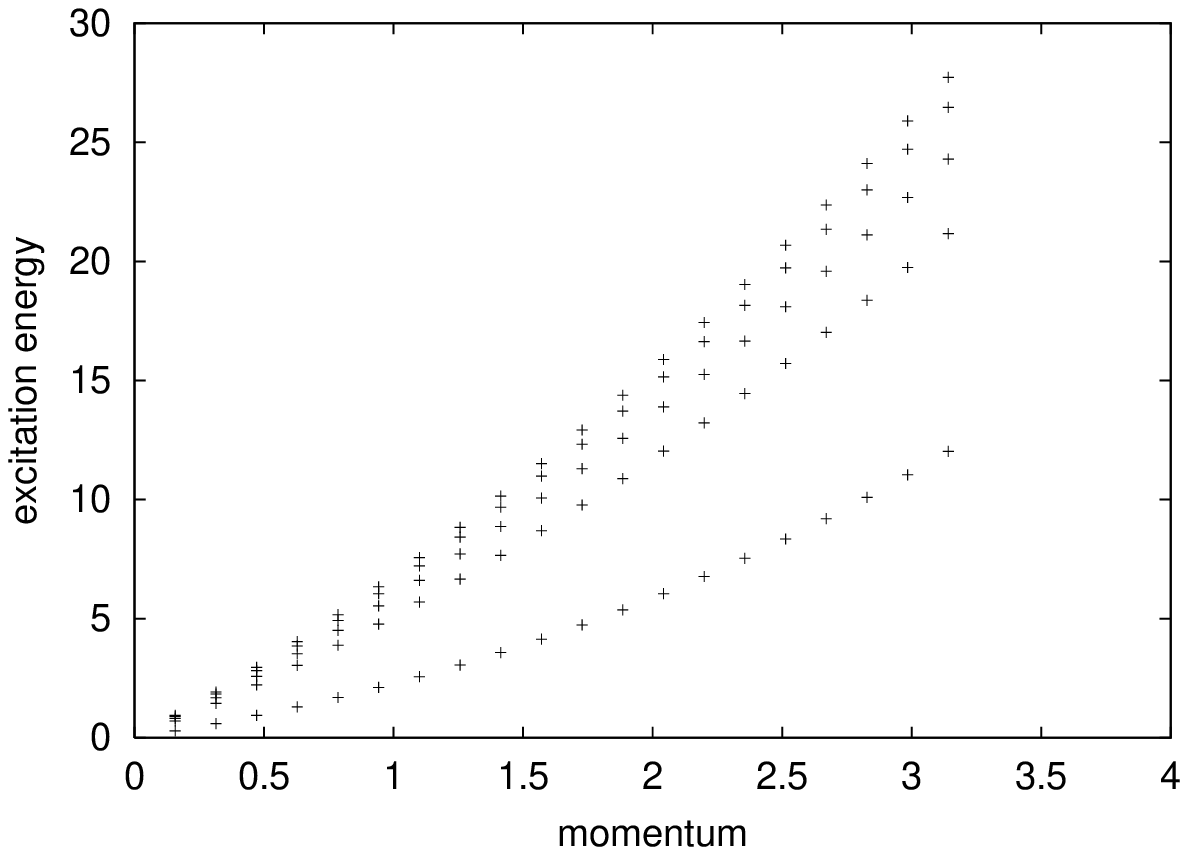}
\caption{The dispersion relation of antiholon excitation for
different coupling constants where the curves from
bottom to top correspond to $c=1, 10, 20, 40, 80$
respectively. Here $N=L=40$.}
\label{FIGURE_DISK}
\end{figure}

\begin{figure}
\setlength\epsfxsize{75mm}
\epsfbox{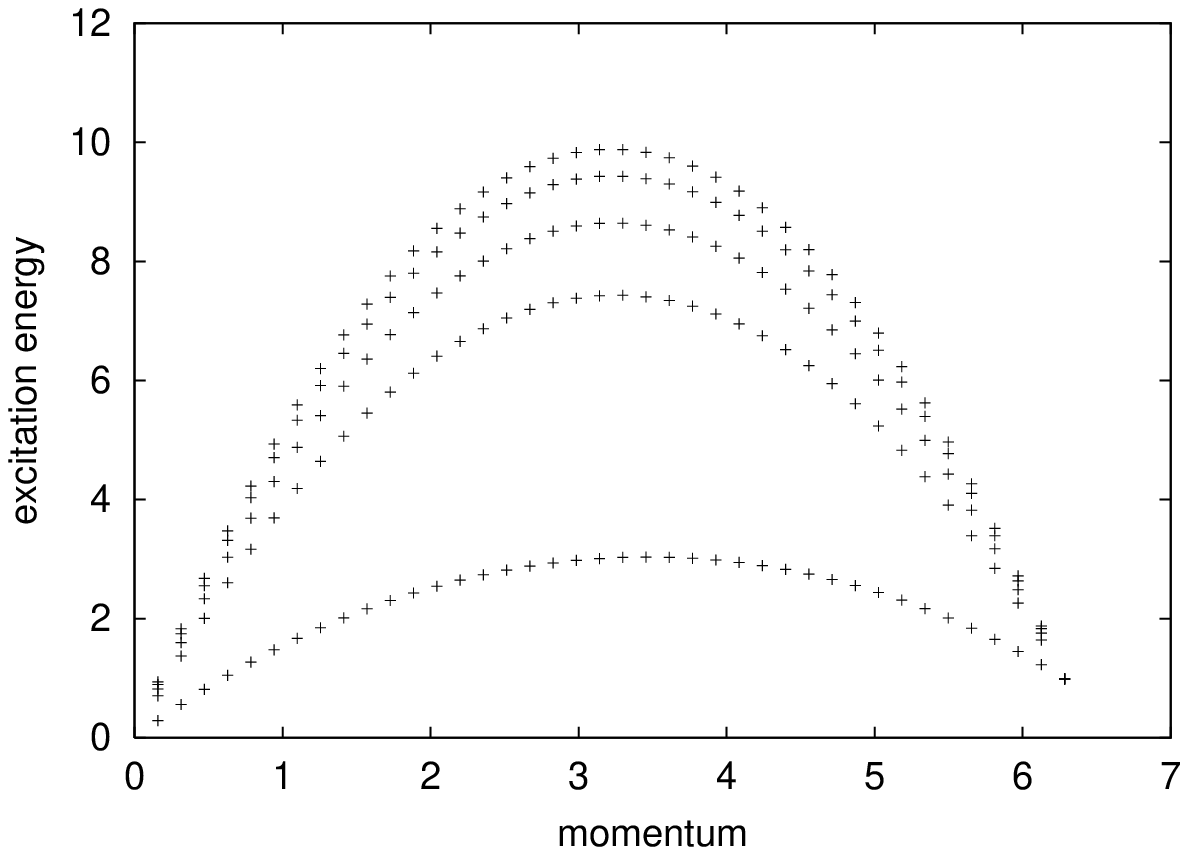}
\caption{The dispersion relation of holon for
different coupling constants where the curves from bottom
to top correspond to $c=1, 10, 20, 40, 80$
respectively. Here $N=L=40$.}
\label{FIGURE_DISH}
\end{figure}

In the configuration of quantum numbers for the ground state
(\ref{eq:GN}), replacing  $I^0_N=(N-1)/2$ by
$I^0_N=(N-1)/2+n,\;n=1,2,\cdots$ and keeping the others unchanged,
we obtain the dispersion relation of antiholon (Fig.
\ref{FIGURE_DISK}) by solving the Bethe ansatz equations
(\ref{eq:logBAE3}) numerically. In a similar way, replacing
$I^0_n,\; n=1,\cdots, N$ of $\{I^0_j\}$ in turn by $(N+1)/2$, we
have the dispersion relation of holon which is shown in Fig.
\ref{FIGURE_DISH}.

In the thermodynamic limit, it is plausible to calculate the
excitation energy by making $\rho(k)=\rho_0(k)+\rho_1(k)/L$ where
$\rho_0(k)$ is the density of the ground state. By creating a hole
inside the quasi Fermi sea $\bar{k}\in [-k_F, k_F]$ and an
additional $k_p>k_F$ outside it, we have the
\begin{eqnarray}
\rho_1(k)+\delta(k-\bar k)=&&\int_{-k_F}^{k_F}dk^\prime
\rho_1(k^\prime)K_2(k-k^\prime) \nonumber \\
&&+K_2(k-k_p).
\end{eqnarray}

The excitation energy consists of two terms $\triangle E=\int k^2
\rho_1(k)dk+k_p^2 =\varepsilon_h({\bar k})+\varepsilon_a(k_p)$.
The holon energy $\varepsilon_h$ and antiholon energy
$\varepsilon_a(k_p)=-\varepsilon_h(k_p)$ are given by,
\begin{eqnarray}
\varepsilon_h({\bar k})=&&-{\bar k}^2
  +\int_{-k_F}^{k_F}k^2\rho_1^h(k,{\bar k})dk, \nonumber \\
\rho_1^h(k,{\bar k})=&&-K_2(k-{\bar k}) \nonumber \\
   &&+\int_{-k_F}^{k_F}K_2(k-k^\prime)\rho_1^h(k^\prime, {\bar k})dk'.
\label{eq:HAE}
\end{eqnarray}

\begin{figure}
\epsfclipoff
\fboxsep=0pt
\setlength{\unitlength}{0.8mm}
\begin{picture}(80,50)(0,0)
\linethickness{1pt}
\epsfysize=45mm
\put(-1,1){{\epsffile{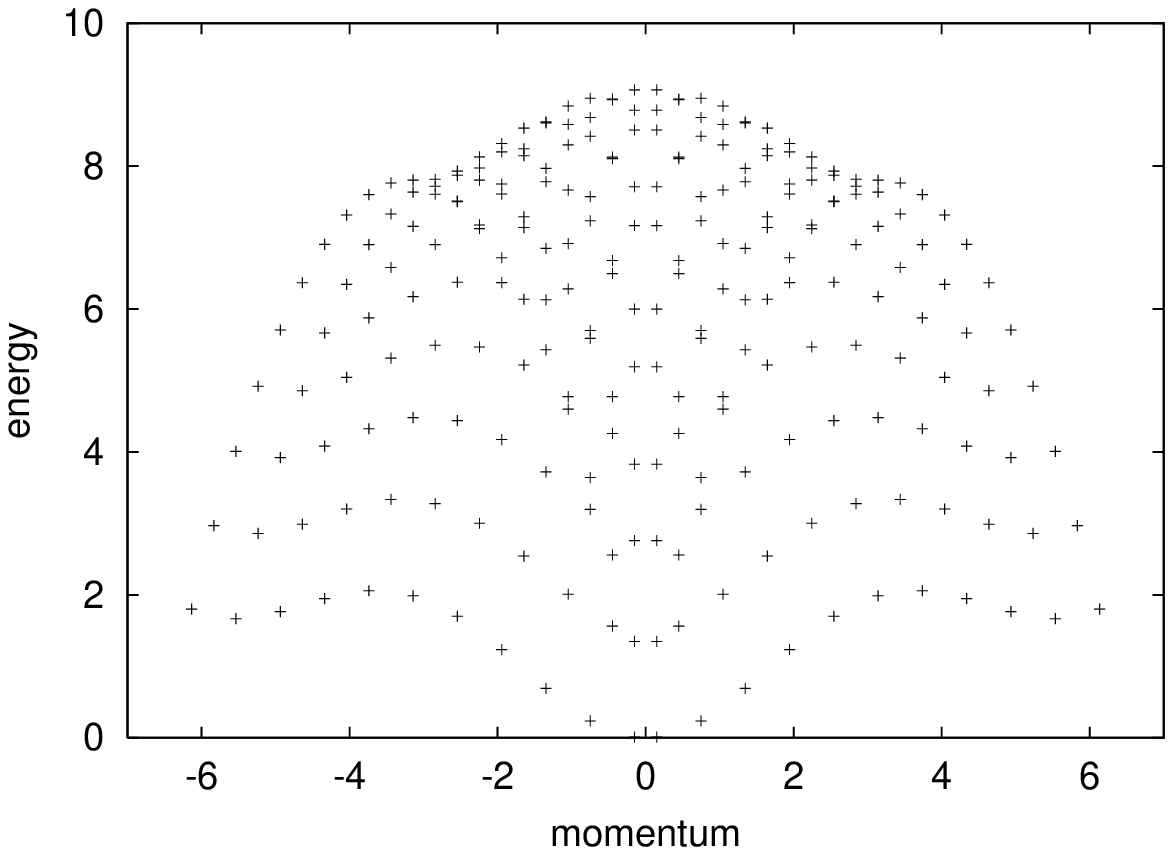}}}
\epsfysize=45mm
\put(90,1){{\epsffile{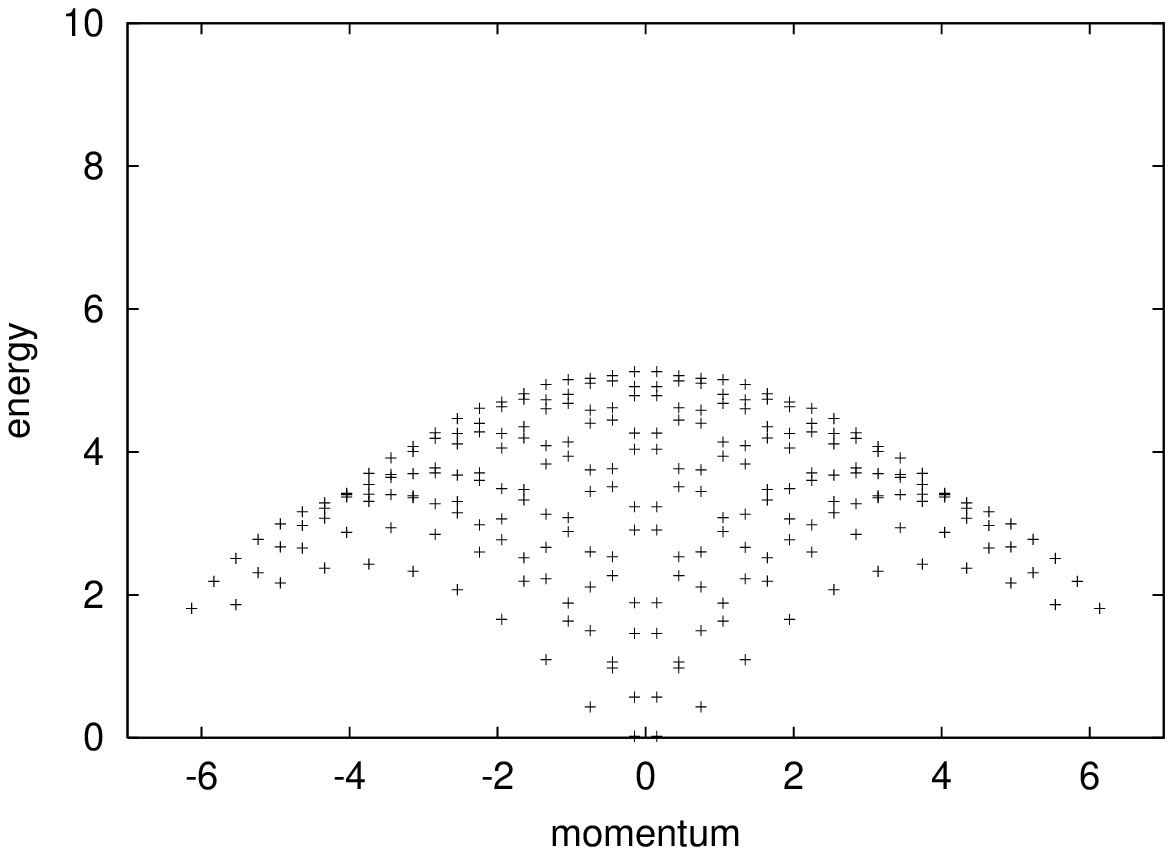}}}
\end{picture}
\vspace{0mm}
\caption{The holon-coloron($\sigma$ type) excitation spectrum
calculated for $N=L=21$ and $c=10$ (left), $c=1$(right).}
\label{FIGURE_HC}
\end{figure}

{\bf Holon-coloron excitation}: Excitations related to the color
sector is characterized by adding $\lambda$ and $\mu$ rapidities
into the system. The simplest excitation of this type is obtained
by considering $M=1$ which is labeled by $(N/2-1,1/2)$. Comparing
to the ground state, the quantum number changes from half-integer
to integer or vice versa. We call this type excitation
$\sigma$-coloron which is regarded as an elementary quasi-particle
of the present model. It's  quantum number takes
\[
I_1=-N/2+\delta_{1,j_1}\;\;\;(1\leq j_1\leq N+1),
\]
\[
I_j=I_{j-1}+1+\delta_{j,j_1}\;\;(j=2,\dots,N),
\]
while $J_1=I_1+m$ $(m=1,2,...N-1)$ so that $I_1<J_1<I_N$. This
produces a $N-1$ multiplets. The excitation spectrum are plotted
in Fig. \ref{FIGURE_HC} for a system of $N=L=21$ with $c=10, 1$
respectively.

Adding an additional $\lambda$ rapidity to the color ferromagnetic
ground state brings about a hole in the $k$-sector. Now we have
two-parameter excitation $\triangle E=\int k^2 \rho_1(k)dk$ where
the $\rho_1(k)$ solves
\begin{eqnarray}
\rho_1(k)+\delta(k-{\bar k})=&&\int_{-k_F}^{k_F} K_2
(k-k^\prime)\rho_1(k^\prime)dk^\prime \nonumber \\
&&-K_1(k-\lambda).
\end{eqnarray}

\begin{figure}
\setlength\epsfxsize{75mm} \epsfbox{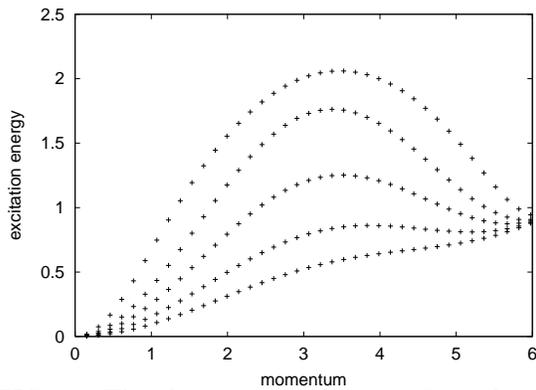} \caption{The
dispersion relation of $\sigma$-coloron for different coupling
constants where the curves from top to bottom correspond to $c=1,
10, 20, 40, 80$ respectively. Here $N=L=41$.} \label{FIGURE_EKC}
\end{figure}

The energy of the holon-coloron excitation consists of two terms
$\triangle E=\varepsilon_h({\bar k})+\varepsilon_c(\lambda)$. The
$\varepsilon_h$ is determined by Eqs. (\ref{eq:HAE}) and the
$\varepsilon_c$ defined by $\varepsilon_c(\lambda)=\int
k^2\rho_1^c(k,\lambda)$ with
\begin{eqnarray}
\rho_1^c(k,\lambda)=&&-K_1(k-\lambda)\nonumber \\
&&+\int_{-k_F}^{k_F}
 K_2(k-k^\prime)\rho_1^c(k^\prime,\lambda)dk^\prime.
\label{eq:Dispcoloron}
\end{eqnarray}
The $\varepsilon_h(\bar{k})$ and $\varepsilon_c(\lambda)$ are
energies of holon and $\sigma$-coloron whose dispersons are shown
in Fig. \ref{FIGURE_DISH} and Fig. \ref{FIGURE_EKC} respectively.

Furthermore, the overall structure for the case of $c=1$ in Fig.
\ref{FIGURE_HC} differs from the case of $c=10$. We interpret the
phenomenon as being due to the fact that the dependence of the
dispersion relation of holon and $\sigma$-coloron on the coupling
constant are different. This feature can be concluded from Fig.
\ref{FIGURE_DISH} and Fig. \ref{FIGURE_EKC}. When $c$ decreases,
the $\varepsilon_h(p)$ decreases while $\varepsilon_c(p)$
increases.

\begin{figure}
\epsfclipoff
\fboxsep=0pt
\setlength{\unitlength}{0.8mm}
\begin{picture}(80,50)(0,0)
\linethickness{1pt}
\epsfysize=45mm
\put(-1,1){{\epsffile{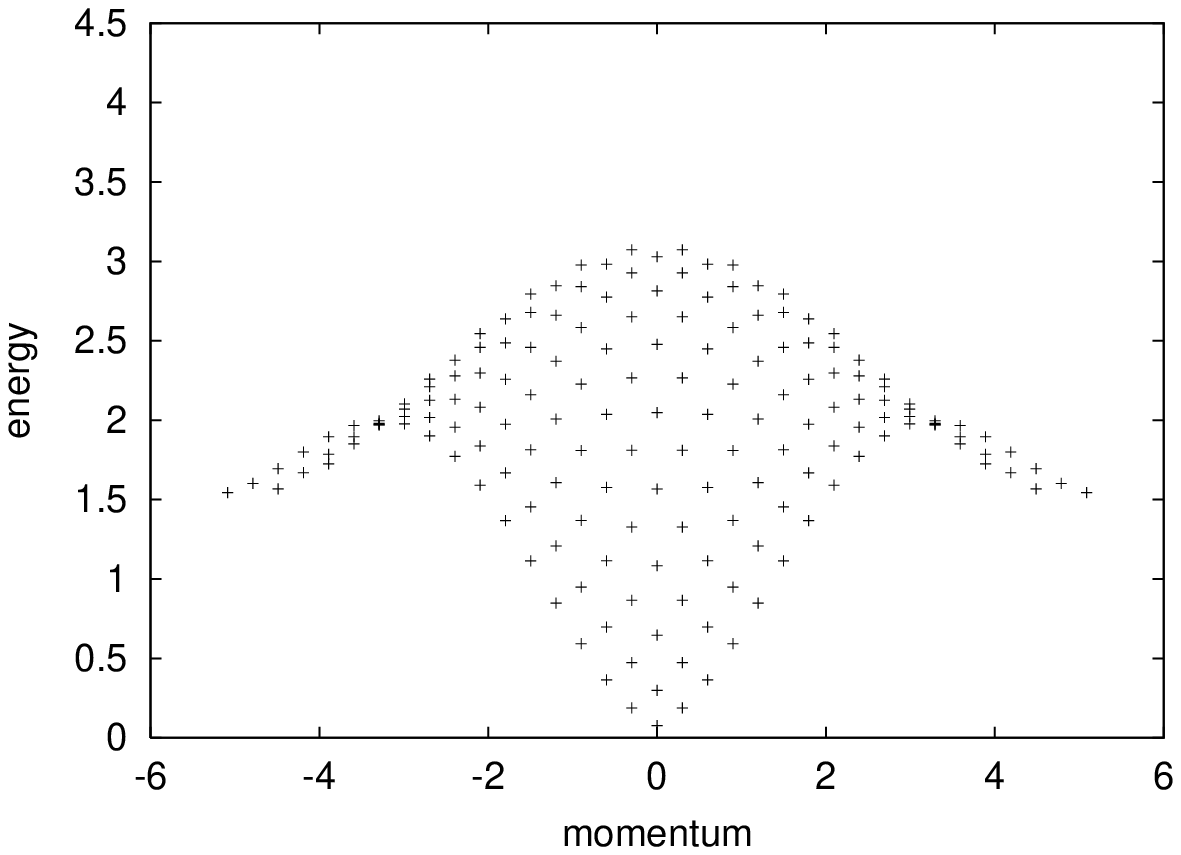}}}
\epsfysize=45mm
\put(90,1){{\epsffile{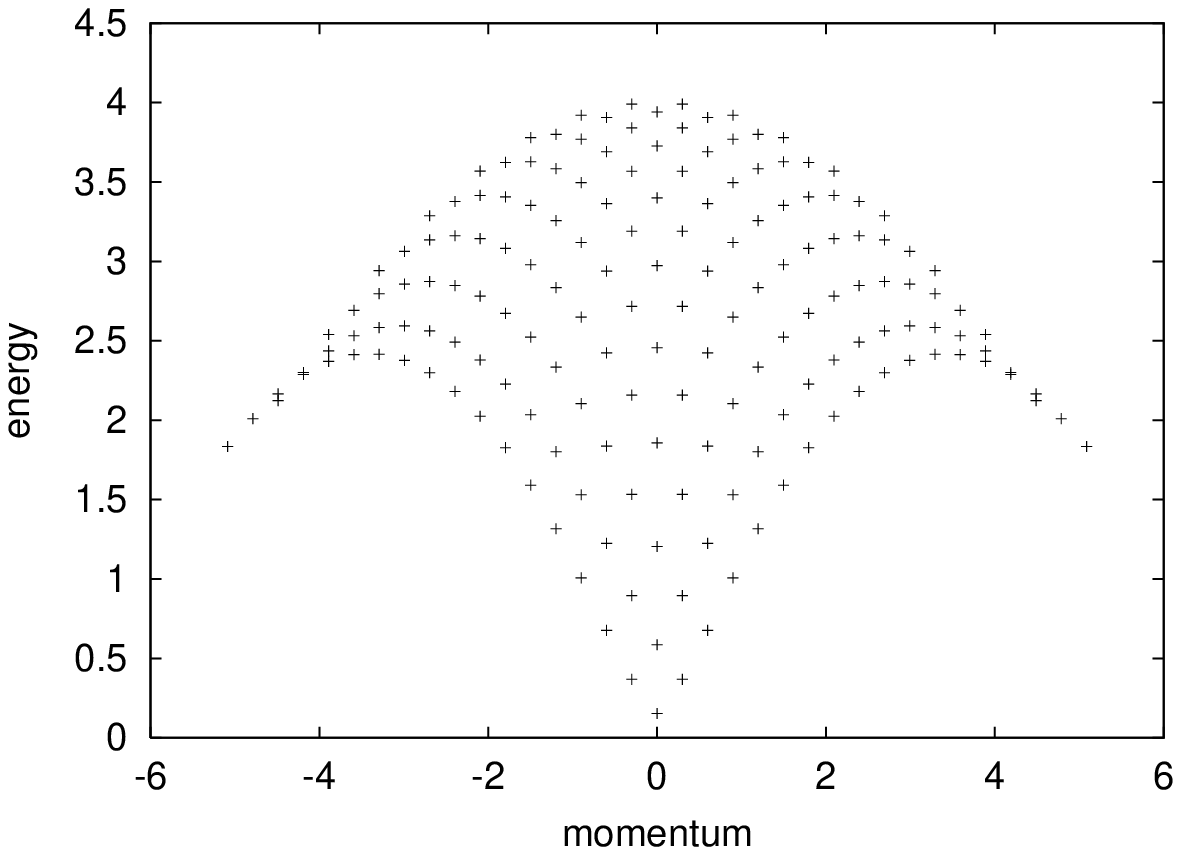}}}
\end{picture}
\vspace{0mm}
\caption{The coloron($\sigma$ type)-coloron($\sigma$ type) excitation spectrum
calculated for $N=L=21$ and $c=10$ (left), $c=1$(right).}
\label{FIGURE_CC}
\end{figure}

{\bf The $\sigma$-type coloron-coloron excitation}: Leaving the
configuration of the ground state $\{I^0_j\}$ unchanged and
changing $M$ from zero to $M=2$, which corresponds to $(N/2-2,1)$,
a two parameters excitation in $\lambda$-sector is characterized
by
\[
-(N-1)/2<J_1<J_2<(N-1)/2.
\]
There are totally $N(N-3)/2$ possible choices for such type of
excitation. In Fig. \ref{FIGURE_CC}, we plot the numerical result
of energy-momentum spectrum for a system with $N=L=21$.

The excitation energy of coloron($\sigma$ type)-coloron ($\sigma$
type) can be calculated by
$\triangle E=\int
k^2\rho_1^c(k,\lambda_1, \lambda_2)dk$,
where
$\rho_1^c(k,\lambda_1, \lambda_2)$ is determined by
\begin{eqnarray}
\rho_1^c(k, \lambda_1, \lambda_2)=&&-K_1(k-\lambda_1)-K_1(k-\lambda_2)
  \nonumber \\
&&+\int_{-k_F}^{k_F}K_2(k-k')\rho_1^c (k',\lambda_1,\lambda_2)dk'.
\end{eqnarray}
It consists of two terms $\triangle
E=\varepsilon_c(\lambda_1)+\varepsilon_c(\lambda_2)$, where the
coloron energy $\varepsilon_c(\lambda)$ has been give in the
passage before Eq. (\ref{eq:Dispcoloron}).

\begin{figure}
\setlength\epsfxsize{75mm} \epsfbox{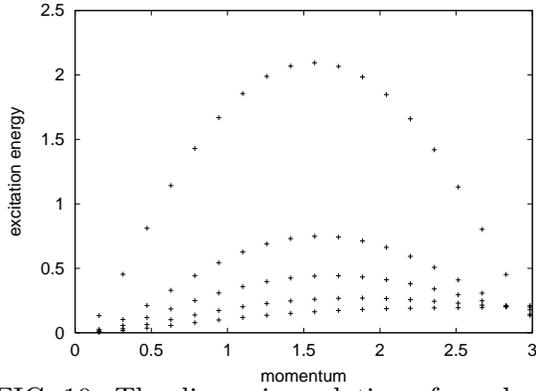} \caption{The
dispersion relation of $\omega$-coloron for different coupling
constants where the curves from top to bottom correspond to $c=1,
10, 20, 40, 80$ respectively. Here $N=L=40$. Zero energy
corresponds to $M=N/2$ ground state.} \label{FIGURE:EKC2}
\end{figure}

{\bf Dispersion relation of $\omega$-coloron}: The fourth possible
excitation involves both the additional quantum number $J$ and
$J'$. For $M=2$ and $M'=1$ there is no range for $J'$ varying, but
for large $M$ the excitation is no more low-lying excitations. So
we only show its dispersion relation which is described by the
following configuration
\begin{eqnarray}
\{I_j\}&&=\{-(N-1)/2,\cdots,(N-1)/2\},\nonumber \\
\{J_\gamma\}&&=\{-M/2,\cdots,(M-2)/2\},\nonumber \\
{J'}_1&&=-M/2+1,\cdots, M/2-1,
\end{eqnarray}
for a given $M$. We plotted the dispersion relation of
$\omega$-coloron in Fig. \ref{FIGURE:EKC2} by varying ${J'}_1$ for
a system with $L=N=40$. The figure has a minimum around $p=\pi$
when $M=N/2$.

Up to now, we have discussed three low-lying excitation energies
and the dispersion relations of four possible elementary
particles, holon, antiholon, $\sigma$-coloron and
$\omega$-coloron. We found those low-lying excitations are gapless
in thermodynamic limit.


\section{Thermodynamics at finite temperature}\label{sec:thermo}
For the ground state (i.e. at zero temperature), the charge
rapidity $k$s are real roots of the Bethe-ansatz equations
(\ref{eq:logBAE3}). For the excited state, however, the $\lambda$
and $\mu$ rapidity can be complex roots\cite{Takahashi,Lai} which
always from a ``bound state'' with several other $\lambda$s. This
arises from the consistency of both hand sides of the Bethe-ansatz
equations\cite{FWoynarovich82} in the limit $L\rightarrow\infty$,
$N\rightarrow\infty$. The n-string rapidity is defined by
\begin{eqnarray}
\Lambda_a^{nj}=\lambda_a^n+(n+1-2j)iu+O(\exp(-\delta N)),\nonumber \\
U_a^{nj}=\mu_a^n+(n+1-2j)iu+O(\exp(-\delta N))
\label{stringdefine}
\end{eqnarray}
where $u=c/2$, $j=1, 2\cdots n$. The total number of $\lambda$ and
$\mu$ are determined by
\begin{equation}
M=\sum_{n=1}^\infty nM_n; \;\;\; M'=\sum_{n=1}^\infty nM'_n.
\end{equation}
where $M_n$ and $M'_n$ denote the number of $\lambda$ n-strings
and $\mu$ n-string respectively. The Eqs. (\ref{eq:logBAE3})
become
\twobeone
\begin{eqnarray}
k_j L=&& 2\pi I_j
    +\sum_l\Theta_1(k_j-k_l)
     +\sum_{an}\Theta_{-n/2}(k_j-\lambda_a^n)
 \nonumber \\
2\pi J_a^n=&&\sum_l\Theta_{-n/2}(\lambda_a^n-k_l)
   +\sum_{bl, t\neq 0}A_{nlt}\Theta_{t/2}(\lambda_a^n-\lambda_b^l)
    +\sum_{clt}B_{nlt}\Theta_{-t/2}(\lambda_a^n-\mu_c^l),
\nonumber \\
2\pi {J'}_a^n=&&\sum_{blt}B_{nlt}\Theta_{t/2}(\mu_a^n-\lambda_b^l)
   +\sum_{cl, t\neq 0}A_{nlt}\Theta_{-t/2}(\mu_a^n-\mu_c^l).
\label{eq:seqular}
\end{eqnarray}
\onebetwo
where
\[
A_{nlt}=\left\{\begin{array}{ll}
1, & {\rm for}\; t=n+l, |n-l|,\\
2, & {\rm for}\; t=n+l-2,\cdots,|n-l|+2,\\
0,  & {\rm  otherwise.}
\end{array}\right.
\]
and
\[
B_{nlt}=\left\{\begin{array}{ll}
1, & {\rm for}\; t=n+l-1, n+l-3,...,|n-l|+1,\\
0,  & {\rm  otherwise.}
\end{array}\right.
\]
and the quantum numbers $\{I_j, J_a^n, {J'}_a^n\}$ label the state
which is no more the ground state. Replacing $k_j, \lambda_a^n,
\mu_a^n$ in Eqs. (\ref{eq:seqular}) by continuous variables $k,
\lambda, \mu$ but keeping the summation still over the solutions
of these roots, we can consider the quantum number $I_j, J_a^n,
{J'}_a^n$ as functions $I(k), J^n(\lambda)$ and $J'^n(\mu)$ given
by Eqs. (\ref{eq:seqular}). Take $I(\lambda)$ as an example, when
$I(k)$ passes through one of the quantum number $I_j$, the
corresponding $k$ is equal to one of the roots $k_j$, so is for
$J^n(\lambda)$ and $J'^n(\mu)$. However, there may exist some
integers or half-integers for which the corresponding
$k$$(\lambda, \mu)$ is not in the set of roots. Such a situation
is conventionally referred as a ``hole". In the thermodynamic
limit, we may introduce the densities of real $k$, $\lambda$
n-string and $\mu$ n-string
\begin{eqnarray}
\rho(k)+\rho^h(k)=&&(1/L)dI(k)/dk,\nonumber \\
\sigma_n(\lambda)+\sigma_n^h(\lambda)=&&(1/L)
  dJ^n(\lambda)/d\lambda, \nonumber \\
\omega_n(\mu)+\omega_n^h(\mu)=&&(1/L)dJ'^n(\mu)/d\mu.
\end{eqnarray}
Then Eqs. (\ref{eq:seqular}) give rise to the following coupled integral
equations,
\begin{eqnarray}
\rho+\rho^h=&&\frac{1}{2\pi}
+\int K_2(k-k^\prime)\rho(k^\prime)dk^\prime \nonumber \\&&
-\sum_n\int K_n(k-\lambda)\sigma_n(\lambda)d\lambda,
\nonumber \\
\sigma_n^h=&&\int K_n(\lambda-k)\rho(k)dk \nonumber \\&&
 -\sum_{lt}A_{nlt}\int K_t(\lambda-\lambda^\prime)
 \sigma_l(\lambda^\prime)d\lambda^\prime \nonumber \\&&
 +\sum_{lt}B_{nlt}\int K_t(\lambda-\mu)\omega_l(\mu)d\mu,
\nonumber \\
\omega_n^h=&&\sum_{lt}B_{nlt}\int K_t(\mu-\lambda)\sigma_l(\lambda)d\lambda
\nonumber \\&&
 -\sum_{lt}A_{nlt}\int K_t(\mu-\mu^\prime)
 \omega_l(\mu^\prime)d\mu^\prime.
\label{eq:density}
\end{eqnarray}
The $\sigma_n$ and $\omega_n$ arising from the definition
(\ref{eq:density}) that occur in the left hand side have been
moved to right hand side by including the $t=0$ term in the
summation. In terms of densities defined above, the total numbers
of $\lambda$ and $\mu$ are given by
\begin{eqnarray}
M/L=&&\sum_n n\int \sigma_n(\lambda)d\lambda, \nonumber \\
M'/L=&&\sum_n n\int \omega_n(\mu)d\mu.
\end{eqnarray}

In the  presence of SU(3) magnetic field $H_1$ and $H_2$, we
can define two type of ``magnetization" whose $z$-components are
\begin{eqnarray}
T^z/L=&&\frac{1}{2}\int \rho(k)dk-\sum_n
n\int\sigma_n(\lambda)d\lambda
   \nonumber \\&&
   +\frac{1}{2}\sum_n n\int \omega_n(\mu)d\mu, \nonumber \\
U^z/L=&&\frac{1}{2}\sum_n n\int \sigma_n(\lambda)d\lambda
  -\sum_n n\int\omega_n(\mu)d\mu.
\end{eqnarray}
Hence the energy contributed by the Zeemann term (\ref{eq:Zee}) is
\begin{equation} E_{Zee}=H_1 T^z + H_2 U^z.
\end{equation}
For given $\rho(k)$, $\rho^h(k)$, $\sigma_n(\lambda)$,
$\sigma_n^h(\lambda)$, $\omega_n(\mu)$ and $\omega_n^h(\mu)$ the
entropy has the form\cite{CNYang69}
\begin{eqnarray}
{\cal S}/L=\int \bigl[(\rho+\rho^h)\ln(\rho+\rho^h)
 -\rho\ln\rho-\rho^h\ln\rho^h \bigr]dk
  \nonumber\\
+\sum_n\int\bigl[(\sigma_n+\sigma_n^h)\ln(\sigma_n+\sigma_n^h)
 -\sigma_n\ln\sigma_n -\sigma_n^h\ln\sigma_n^h\bigr]d\lambda
   \nonumber \\
+\sum_n\int\bigl[(\omega_n+\omega_n^h)\ln(\omega_n+\omega_n^h)
-\omega_n\ln\omega_n -\omega_n^h\ln\omega_n^h\bigr]d\mu.
\end{eqnarray}
where the Boltzmann constant is set to unit.

At finite temperature, the thermal equilibrium is obtained by
minimizing the free energy $F=E-E_{Zee}-T{\cal S}-\mu N$ where
$\mu$ is the chemical potential and $\cal{S}$ is the entropy of
the system. Making use of the relations derived from Eqs.
(\ref{eq:density}) \twobeone
\begin{eqnarray}
\delta \rho^h=&&-\delta\rho +\int K_2(k-k^\prime)\delta\rho dk'
   -\sum_n\int K_n(k-\lambda)
   \delta\sigma_n d\lambda,
     \nonumber \\
\delta\sigma_n^h=&&\int K_n(\lambda-k)\delta\rho dk
  -\sum_{lt}A_{nlt}\int K_t(\lambda-\lambda^\prime)
   \delta\sigma_l(\lambda^\prime)d\lambda^\prime
  +\sum_{lt}B_{nlt}\int K_t(\lambda-\mu)
   \delta\omega_l(\mu)d\mu,
    \nonumber \\
\delta\omega_n^h=&&\sum_{lt}B_{nlt}\int K_t(\mu-\lambda)
   \delta\sigma_l(\lambda)d\lambda
   -\sum_{lt}A_{nlt}\int K_t(\mu-\mu')
   \delta\omega_l(\mu^\prime)d\mu',
\end{eqnarray}
and define
\begin{eqnarray}
\frac{\rho^h(k)}{\rho(k)}=&&\kappa(k)=e^{\epsilon(k)/T},\;\;\;
\frac{\sigma_n^h(\lambda)}{\sigma_n(\lambda)}
=\eta_n(\lambda)=e^{\zeta_n(\lambda)/T},\;\;\;
\frac{\omega_n^h(\mu)}{\omega_n(\mu)}
=\Delta_n(\mu)=e^{\xi_n(\mu)/T}.
\end{eqnarray}

We obtain the following conditions from the
minimum condition $\delta F=0$, namely
\begin{eqnarray}
\epsilon(k)=&&k^2-\mu-H_1/2
-T\int K_2(k-k^\prime)
 \ln(1+e^{-\epsilon(k^\prime)/T})dk'
-T\sum_n\int K_n(k-\lambda)
\ln[1+e^{-\zeta_n(\lambda)/T}]d\lambda
  \nonumber \\
\zeta_n(\lambda)=&&n(2H_1-H_2)/2
+T\int K_n(\lambda-k)\ln[1+e^{-\epsilon(k)/T}]dk
+T\sum_{l,t\neq 0}A_{nlt}\int K_t(\lambda-\lambda^\prime)
   \ln[1+e^{-\zeta_l(\lambda^\prime)/T}]d\lambda^\prime
    \nonumber \\
&&-T\sum_{lt}B_{nlt}\int K_t(\lambda-\mu)
   \ln[1+e^{-\xi_l(\mu)}]d\mu,
    \nonumber \\
\xi_n(\mu)=&&n(2H_2-H_1)/2
  -T\sum_{lt}B_{nlt}\int K_t(\mu-\lambda)
   \ln[1+e^{-\zeta_l(\lambda)}]d\lambda
    \nonumber \\
&&+T\sum_{l,t\neq 0}A_{nlt}\int K_t(\mu-\mu^\prime)
   \ln[1+e^{-\xi_l(\mu^\prime)/T}]d\mu^\prime
\label{thermalequations}
\end{eqnarray}
A more useful version of Eqs. (\ref{thermalequations}) is the
recursive scheme which is a revised version of Gaudin-Takahashi
equations, as obtained by Fourier transform.
\begin{eqnarray}
T\ln\kappa=&&k^2-\mu-H_1/2-T K_2(k)*\ln[1+\kappa^{-1}]
 -T\sum_n K_n(k)*\ln[1+\eta_n^{-1}],\nonumber \\
\ln\eta_1=&&\frac{1}{4u}{\rm sech}(\pi\lambda/2u)*
  \ln[(1+\kappa^{-1})(1+\eta_2)/(1+\Delta_1^{-1})], \nonumber \\
\ln\eta_n=&&\frac{1}{4u}{\rm sech}(\pi\lambda/2u)*
  \ln[(1+\eta_{n-1})(1+\eta_{n+1})/(1+\Delta_n^{-1})], \nonumber \\
\ln\Delta_1=&&\frac{1}{4u}{\rm sech}(\pi\lambda/2u)*
  \ln[(1+\Delta_2)/(1+\eta_1^{-1})], \nonumber \\
\ln\Delta_n=&&\frac{1}{4u}{\rm sech}(\pi\lambda/2u)*
  \ln[(1+\Delta_{n-1})(1+\Delta_{n+1})/(1+\eta_n^{-1})].
\label{thermalequations2}
\end{eqnarray}
\onebetwo
where $*$ denotes a convolution. And these equations are
complete by the asymptotic conditions
\begin{eqnarray}
\lim_{n\rightarrow\infty}[\ln\eta_n/n]=&&(2H_1-H_2)/2T,\nonumber \\
\lim_{n\rightarrow\infty}[\ln\Delta_n/n]=&&(2H_2-H_1)/2T.
\label{eq:asycon}
\end{eqnarray}
Finally, we obtain the Helmholtz free energy $F=E-T{\cal S}$:
\begin{eqnarray}
F=&&\mu N-\frac{LT}{2\pi}\int\ln[1+e^{-\epsilon}]dk,
\label{GFE}
\end{eqnarray}
and the pressure of the system
\begin{equation}
P=-\frac{\partial F}{\partial L}=\frac{T}{2\pi}\int\ln[1+e^{-\epsilon}]dk,
\end{equation}
which is formally the same as Yang and Yang's expression but the
equation which $\epsilon$ fulfills is different.

\section{Special Cases}\label{sec:spec}

In general, the free energy can be calculated by using formula
(\ref{GFE}), where $\epsilon(k)$ and $\zeta_n(\lambda)$ are
determined from Eqs. (\ref{thermalequations}) which can be solved
by iteration. In the following we will consider some special cases
respectively as explicit results are obtainable in those cases.

\subsection{Zero temperature limit}
The state at zero temperature is the ground state. When
$T\rightarrow 0$, the first equation of Eqs.
(\ref{thermalequations}) becomes
\begin{eqnarray}
\epsilon(k)=&&k^2-\mu-H_1/2
+\int K_2(k-k^\prime)
  \epsilon(k^\prime)dk^\prime \nonumber \\
&&+\sum_n\int K_n(k-\lambda)
\zeta_n(\lambda)d\lambda
\label{eq:Dress}
\end{eqnarray}
Then the Fermi surface is determined by $\epsilon(k_F)=0$. Since
there is no hole under Fermi surface, we can take the ratio
$\kappa=\rho^h/\rho$ as zero when $k\in[-k_F, k_F]$. As a result,
it is easy to see from Eqs. (\ref{thermalequations2}) that
$\eta_n=\Delta_n\rightarrow\infty$. That is $M=M'=0$, the state is
color ferromagnetic state. This is consistent with the conclusion
obtained in Sec. \ref{sec:low}. Then Eq. (\ref{eq:Dress}) can be
rewritten as
\begin{equation}
\epsilon_0(k)=k^2-\mu-H_1/2 +\int_{-k_F}^{k_F} K_2(k-k')
  \epsilon_0(k^\prime)dk',
\end{equation}
which gives the solution of dressed energy\cite{Frahm}, and the
ground-state energy can be given in terms of $\epsilon_0$
\begin{equation}
E_0/L=\frac{1}{2\pi}\int_{-k_F}^{k_F}\epsilon_0(k)dk.
\end{equation}
whose dependence on the coupling constant is shown in Fig.
(\ref{FIGURE:EKC2}).

\subsection{High temperature limit}
In the high temperature limit $T\rightarrow \infty$,
we can assume that all functions $\eta_n(\lambda)$ and $\Delta_n(\mu)$
are independent of their corresponding parameter.
Due to $\lim_{u\rightarrow 0}\frac{1}{2u}{\rm sech}(\frac{\pi\lambda}{2u})=\delta(\lambda)$,
Eqs. (\ref{thermalequations2}) can be written as follows,
\begin{eqnarray}
\eta_1^2=&&(1+\kappa^{-1})(1+\eta_2)/(1+\Delta_1^{-1}),\nonumber \\
\eta_n^2=&&(1+\eta_{n-1})(1+\eta_{n+1})/(1+\Delta_n^{-1}), \nonumber\\
\Delta_1^2=&&(1+\Delta_2)/(1+\eta_1^{-1}),\nonumber \\
\Delta_n^2=&&(1+\Delta_{n-1})(1+\Delta_{n+1})/(1+\eta_n^{-1}).
\label{eq:dgfkkgdd}
\end{eqnarray}
with the asymptotic conditions Eqs. (\ref{eq:asycon}).

Performing  the Fourier transform to Eqs. (\ref{eq:density}), we
get the solution of the densities of $\lambda$ n-strings,
\begin{eqnarray}
\sigma_1+\sigma_1^h=&&\frac{1}{4u}{\rm sech}[\pi\lambda/2u]*[\rho+\sigma_2^h+\omega_1],
\nonumber \\
\sigma_n+\sigma_n^h=&&\frac{1}{4u}{\rm sech}[\pi\lambda/2u]*
 [\sigma_{n-1}^h+\sigma_{n+1}^h+\omega_n], \nonumber \\
 \omega_1+\omega_1^h=&&\frac{1}{4u}{\rm sech}[\pi\lambda/2u]*[\omega_2^h+\sigma_1]
 \nonumber \\
 \omega_n+\omega_n^h=&&\frac{1}{4u}{\rm sech}[\pi\lambda/2u]*
 [\omega_{n-1}^h+\omega_{n+1}^h+\sigma_n].
\end{eqnarray}
If we assume that $\sigma_n$, $\sigma^h_n$ and $\omega_n, \omega_n^h$
are independent of
$\lambda$  and $\mu$ respectively, or let $u=0$, we have the following relation
\begin{eqnarray}
\sum_n n\sigma_n=&&\frac{\rho}{2}+\frac{1}{2}\sum_n n\omega_n+
-\frac{n_m+1}{2}\sigma_{n_m}e^{n_m\Omega_1/T},
\nonumber\\
\sum_n n\omega_n=&&\frac{1}{2}\sum_n n\sigma_n
-\frac{n_{m'}+1}{2}\sigma_{n_m}e^{n_{m'} \Omega_2/T}
\label{eq:MML}
\end{eqnarray}
where $n_m$ and $n_{m'}$ are the maximal length of $\lambda$
string and $\mu$ string respectively, and $\Omega_1=2H_1-H_2,
\Omega_2=2H_2-H_1$. In the absence of external field $H_1$ and
$H_2$, it is easy to obtain $M'=M/2=N/3$ which means there are
$N/3$ particles in each internal state. Then we can also infer
that the contribution of the internal degree of freedom to the
entropy per site must be ${\cal S}=\ln 3$ which follows from the
fact that the internal degree of freedom per particle is three.

If the external field is small, expanding Eq. (\ref{eq:MML}) for
small field  and integrating the equation over $\lambda$ and
$\mu$, we get the SU(3) magnetization of the model,
\begin{eqnarray}
\frac{T^z}{L}=\frac{M_m}{2L}\left[1+\frac{n_m
 \Omega_1}{T}+\frac{1}{2}\left(\frac{n_m\Omega_1}{T}\right)^2+\cdots\right]
 \nonumber \\
\frac{U^z}{L}=\frac{M'_m}{2L}\left[1+\frac{n_{m'}\Omega_2}{T}+\frac{1}{2}
\left(\frac{n_{m'}\Omega_2}{T}\right)^2+\cdots\right].
\label{eq:MAGNETIZATION}
\end{eqnarray}
where $M_{m}$ and $M_{m'}$ are the total number of rapidities in
$\lambda$ $n_m$-strings and $\mu$ $n_{m'}$-strings respectively.
The first term in the parentheses of both equations arises from
self-magnetization, while the others are contributed by external
field. Eq. (\ref{eq:MAGNETIZATION}) indicates that the magnetic
property of the model in high temperature regime is dominated by
Curie's law $\chi\propto 1/T$.

\subsection{The strong coupling limit}

For $u\rightarrow \infty$, $K_n(k)$ goes to zero, from Eqs.
(\ref{thermalequations}) we have
\begin{equation}
\epsilon=k^2-\mu.
\end{equation}
where the external field is set to unit. The $k$-sector are
completely decoupled with $\lambda$ and $\mu$-sectors. At
arbitrary temperature, the solutions for the $\eta_n, \Delta_n$
are independent of parameters $\lambda$ and $\mu$ respectively,
which gives rise to Eq. (\ref{eq:dgfkkgdd}) The free energy of the
system defined by Eq. (\ref{GFE}) can be solved by integration by
part,
\begin{equation}
F/L=\mu D-\frac{2}{\pi}
  \Bigl[\frac{1}{3}\mu^{3/2}+\frac{T^2\pi^2}{24\mu^{1/2}}\Bigr]
\end{equation}
where the external field is set to zero.

We are not able to deduce the specific heat directly from the free
energy obtained above because the chemical potential is a function
of temperature. From Eqs. (\ref{eq:density}), the density of
charge rapidity has the form
\begin{equation}
\rho=\frac{1}{2\pi} \frac{1}{1+e^{(k^2-\mu)/T}}
\end{equation}
Integrating the charge density over $k$ space with the condition
Eq. (\ref{eq:density222}), we have an explicit expression of the
chemical potential,
\begin{equation}
\mu=\mu_0\Bigl[1-\frac{\pi^2 T^2}{24\mu_0^2}\Bigr]^{-2}
\end{equation}
where $\mu_0=\pi^2 D^2$ which denotes $\mu$ at zero temperature.
Then the free energy becomes
\begin{equation}
F/L=\mu_0 D\Bigl[1+\frac{\pi^2T^2}{12\mu_0^2}\Bigr]
  -\frac{2}{3\pi}
  \mu_0^{3/2}\Bigl[1+\frac{\pi^2T^2}{4\mu_0^2}\Bigr].
\end{equation}
The free energy for the SU(3) invariant spin chain also has a
$T^2$ dependence\cite{Tsvelik}.

Since in thermodynamics $S=-\partial F/\partial T$ and
$C_v=T\partial S/\partial T$, we find the specific heat at low
temperature is Fermi-liquid like,
\begin{equation}
S=C_v=\frac{T}{6D}.
\end{equation}
It is the same as the result of one-component case, since for the
strong coupling limit the color degree of freedom and the charge
degree of freedom are decoupled completely, the contribution of
color degree of freedom to the free energy vanishes.

\section{Conclusions}\label{conclusions}

In this paper, we have solved one dimensional SU(3) bosons with
$\delta$-function interaction by means of coordinate Bethe-ansatz
method. On the basis of Bethe-ansatz equations we first discussed
the ground state of the Bose system and found that the ground
state is a color ferromagnetic state which differs from the SU(3)
Fermi system greatly. The configuration of quantum numbers for the
ground state was given explicitly. The low-lying excitations were
discussed extensively by both analytical and numerical methods.
The energy-momentum spectra for three type excitations:
holon-antiholon, holon-coloron($\sigma$ type) and coloron($\sigma$
type)-coloron($\sigma$ type) were plotted for $c=10$ and $c=1$. We
also discussed the dispersion relations of four elementary
quasi-particles.

The thermodynamics of the system were studied by using the
strategy of TBA. A revised version of Gaudin-Takahashi equations
were obtained by minimizing the free energy at finite temperature.
We found the magnetic property of the system at high temperature
regime is dominated by Curie's law, and for the case of strong
coupling  the system possess the properties of Fermi-liquid like
and its specific heat is a linear function of $T$ at low
temperature.

\section*{Acknowledgement}
This work is supported by trans-century projects and Cheung Kong
projects of China Education Ministry.

\twobeone

\end{document}